\documentclass{article}
\usepackage{authblk}
\usepackage{graphicx}
\usepackage{amsmath}
\usepackage{amssymb}
\usepackage{booktabs}
\usepackage{geometry}
\usepackage{caption}
\usepackage{subcaption}
\usepackage{hyperref}
\usepackage{multicol}
\usepackage{array}

\usepackage[backend=biber, style=numeric]{biblatex}
\usepackage{float}  
\title{A DNA Methylation Classification Model Predicts Organ and Disease Site}

\author[1*]{Keng-Jung Lee}
\author[2*]{Dharanya Sampath}
\author[3*]{Konstantinos Mavrommatis}
\affil[1]{Carnegie Mellon University, Department of Biomedical Engineering, Pittsburgh, Pennsylvania, United States}
\affil[2]{Bristol Myers Squibb, 1000 Sierra Point Pkwy, Brisbane, California 94005, United States}
\affil[3]{Bristol Myers Squibb, Boudry, Route de Perreux 1, Switzerland}
\affil[*]{For correspondence: kengjunglee@cmu.edu, Dharanya.Sampath@bms.com, Konstantinos.Mavrommatis@bms.com}
\date{}

\begin{document}
\maketitle

\begin{abstract}
Cell-free DNA (cfDNA) analysis has emerged as a powerful, minimally invasive tool for monitoring disease progression, treatment response, and early detection. However, one of the major hurdles in cfDNA-based diagnostics is accurately determining the tissue of origin, especially in complex or het erogeneous disease contexts. To address this, we developed a robust machine learning framework that leverages tissue-specific DNA methylation signatures to classify both tissue and disease origin from cfDNA data. Our model integrates DNA methylation datasets generated across diverse epigenomic platforms, including Whole Genome Bisulfite Sequencing (WGBS), Illumina Infinium Bead Arrays, and Enzymatic Methyl-seq (EM-seq). To account for platform variability and sparsity in methylation data, we applied imputation strategies and harmonized CpG features to enable cross-platform learning. Dimensionality reduction confirmed clear tissue-specific clustering of methylation profiles, and model training using a random forest classifier achieved consistent performance, with classification accuracies ranging from 0.75 to 0.8 across test sets and platforms. Importantly, our model was able to distinguish clinically relevant tissues such as inflamed synovium and peripheral blood mononuclear cells (PBMCs) in arthritis patients and successfully deconvoluted synthetic cfDNA mixtures that mimic real-world liquid biopsy samples. The predicted probabilities of tissue origin closely correlated with the true proportions in these mixtures, suggesting our model’s potential utility for both qualitative classification and quantitative tissue composition inference. Together, these results highlight the feasibility of using cross-platform methylation data and machine learning to develop scalable, generalizable cfDNA-based diagnostics. Our approach provides a foundation for future integration of disease-specific epigenetic features informing clinical decision-making in precision medicine applications.

\end{abstract}

\section{Introduction}
Tissue biopsies remain the cornerstone of clinical diagnostics and treatment response assessment, particularly in oncology, where histopathologic examination provides definitive insights into disease type, stage, and progression \cite{hoshino_extracellular_2020, zhang_molecular_2020}. Despite their diagnostic value, tissue biopsies are inherently invasive, associated with procedural risks and patient discomfort, and are often impractical in anatomically inaccessible or medically fragile contexts \cite{zhang_molecular_2020}. Furthermore, they offer only a static snapshot of an evolving disease process and are vulnerable to sampling bias due to tumor heterogeneity and spatial variability. While radiographic imaging can complement biopsies, it lacks the molecular resolution required for early detection or granular disease characterization, especially in cases of minimal residual disease or micrometastatic spread\cite{hoshino_extracellular_2020, zhang_molecular_2020}.\\\\
To address these limitations, liquid biopsy has emerged as a promising, minimally invasive diagnostic alternative \cite{luo_liquid_nodate}. By enabling real-time, repeatable access to circulating biomarkers in biofluids (primarily blood biopsies) offer the potential to detect disease earlier, monitor dynamic treatment responses, and non-invasively capture molecular changes across disease progression. Biomarkers such as circulating tumor cells (CTCs), extracellular vesicles (EVs), proteins, and various nucleic acids are increasingly used to inform clinical decision-making. Among these, cell-free DNA (cfDNA) stands out due to its accessibility, abundance, and rich molecular content reflecting both genetic and epigenetic alterations \cite{luo_liquid_nodate}.\\\\
cfDNA consists of short, double-stranded DNA fragments, typically 150–200 base pairs in length, that are released into circulation during cellular turnover via apoptosis, necrosis, or active secretion\cite{luo_liquid_nodate}. Although cfDNA concentrations are low in healthy individuals (usually below 10 ng/mL), they rise substantially in pathological states including cancer, autoimmune disorders, trauma, and cardiovascular injury\cite{luo_liquid_nodate}. Beyond quantitative changes, cfDNA retains a wealth of molecular information, including somatic mutations, chromatin accessibility, and critically DNA methylation patterns\cite{luo_liquid_nodate}.\\\\
DNA methylation is a stable and cell type-specific epigenetic modification that plays a fundamental role in regulating gene expression, chromatin structure, and maintaining cell identity\cite{loyfer_dna_2023}. Because methylation patterns are highly tissue-specific and conserved across physiological and pathological states, they offer a powerful means of tracing the tissue origin of cfDNA. Unlike mutations, which can be rare or heterogeneous across tumors, methylation signals provide consistent and discriminatory features ideal for classification tasks, especially in machine learning applications.\\\\
In this study, we present a platform-agnostic machine learning framework that classifies cfDNA based on its DNA methylation profile to infer both tissue-of-origin and disease state. We first leveraged a comprehensive methylation atlas of human cell and tissue types to train and validate our classification model, achieving high accuracy across a range of organs. We further demonstrated that the model can distinguish between inflamed and non-inflamed tissue contexts, such as synovium versus PBMCs in Rheumatoid Arthritis and Osteoarthritis, with perfect accuracy (ROC AUC = 1.0). Finally, we validated the model's deconvolution capabilities using in silico mixtures of cfDNA from different tissues. The predicted probabilities correlated with input proportions, confirming the model’s quantitative resolution and potential application in complex clinical scenarios such as cancer, autoimmunity, or organ injury.\\\\
Together, our findings highlight the feasibility of using DNA methylation-based machine learning for cfDNA classification and deconvolution. This approach enables high-resolution, non-invasive inference of tissue pathology and origin, paving the way for next-generation precision diagnostics in oncology, immunology, and beyond.
\section{Results}
\subsection{Comparison of machine learning models}
To achieve accurate deconvolution of cfDNA based on DNA methylation, we first established a comprehensive reference map of tissue-specific methylation patterns. We utilized a publicly available methylation atlas comprising 223 cell types and subtypes derived from diverse human organs, profiled by Whole Genome Bisulfite Sequencing (WGBS)\textsuperscript{1}. For each sample, methylation levels at CpG sites were estimated using a dynamic programming framework that modeled local methylation probabilities with a Bernoulli distribution, enabling reliable segmentation of the genome into homogeneously methylated regions. To handle sparsity and missing values inherent in high-throughput methylation datasets, we employed a K-nearest neighbor (KNN) imputation strategy, producing a dense matrix suitable for downstream machine learning.\\\\
To evaluate whether methylation profiles carry distinct and consistent tissue-of-origin signals, we performed dimensionality reduction using Uniform Manifold Approximation and Projection (UMAP). As illustrated in Figure 1a, the resulting embedding revealed clear and coherent clustering of samples according to their annotated tissue and cell type. Organ-level groupings such as muscle, liver, lung, heart, blood, and brain were readily distinguishable, with notable resolution even among closely related epithelial or endothelial subtypes. These results confirm the feasibility of using DNA methylation as a robust molecular fingerprint for tissue classification.\\\\
To train and assess a predictive model, we split the methylation dataset into 70\% training and 30\% testing subsets. Within the training data, we performed 10-fold cross-validation to optimize model hyperparameters and guard against overfitting. Among several machine learning algorithms tested, the Random Forest classifier consistently outperformed others, achieving a testing accuracy of 0.82, demonstrating strong generalization across diverse tissue types.\\\\
The classifier’s performance is summarized in Figure 1b, which shows a normalized confusion matrix across all tested samples. The high diagonal dominance indicates accurate tissue-of-origin prediction for most classes, with minimal confusion among biologically similar tissues. Some expected overlap appeared between closely related tissues (e.g., endodermal-derived epithelial subtypes), yet the majority of samples were correctly classified.\\\\
We further visualized the model’s output probabilities for each test sample in Figure 1c, which maps the predicted probability distribution across all possible tissue labels. The strong confidence scores along the correct labels and minimal background probabilities elsewhere support the discriminative power of methylation-based classification. Importantly, this probability-based output format provides a basis for quantitative tissue deconvolution in mixed-source cfDNA, laying the groundwork for its application in complex clinical settings such as cancer, inflammation, or transplant monitoring.\\\\
Together, these results demonstrate that high-dimensional methylation data—when processed with appropriate statistical and machine learning methods—can effectively classify tissue-of-origin with high accuracy, robustness, and interpretability.
\begin{table}[H]
\centering
\caption{Comparison of model performance on training and testing datasets}
\label{tab:model_performance}
\begin{tabular}{>{\bfseries}l c c}
\toprule
Model & Training Accuracy (70\% data) & Testing Accuracy (30\% untouched data) \\
\midrule
K-Nearest Neighbors & 0.69 & 0.23 \\
Support Vector Machine & 0.82 & 0.6 \\
Random Forest & 1 & 0.82 \\
\bottomrule
\end{tabular}
\end{table}
\subsection{Disease Detection Studies}
To evaluate whether disease-associated DNA methylation alterations are sufficient to distinguish inflamed from non-inflamed tissues, we investigated methylation signatures derived from synovial biopsies and peripheral blood mononuclear cells (PBMCs) in patients with Rheumatoid Arthritis (RA) and Osteoarthritis (OA). These two tissue types represent critical compartments in inflammatory joint disease: PBMCs reflect systemic immune status, while synovium captures the local inflammatory microenvironment. Our goal was to determine whether these disease-relevant tissues exhibit distinct epigenetic profiles that can be leveraged for classification, which is a crucial step toward enabling cfDNA-based diagnostics for autoimmune and inflammatory diseases.\\\\
Using Whole Genome Bisulfite Sequencing (WGBS) data, we processed and extracted methylation signatures from both tissue types. We then applied Uniform Manifold Approximation and Projection (UMAP) to project high-dimensional methylation features into a two-dimensional space. As shown in Figure 2a, PBMC and synovial samples clustered into two clearly separable groups, demonstrating that disease-associated methylation differences are sufficiently pronounced to enable unsupervised separation of tissue origin—even among samples derived from the same individuals and disease backgrounds. Notably, this separation was maintained across both RA and OA cohorts, suggesting that the methylation distinctions are robust and generalizable.\\\\
To further quantify the classification potential of these signatures, we trained a Random Forest classifier to distinguish synovial tissue from PBMCs using methylation features as input. The classifier achieved perfect performance, with an area under the receiver operating characteristic curve (ROC AUC) of 1.0 for both tissue types (shown in Figure 2b for synovium and Figure 2c for PBMCs). These results confirm that inflammation-related epigenetic remodeling in RA and OA leaves a detectable and classifiable imprint on the DNA methylation landscape.\\\\
Together, these findings demonstrate the strong potential for using machine learning-based analysis of DNA methylation to not only infer tissue-of-origin but also detect disease-associated immune activation and local inflammation. This work provides a foundational proof-of-concept for applying cfDNA-based epigenomic deconvolution in autoimmune diseases, with implications for early detection, disease monitoring, and personalized therapeutic strategies.\\\\
\begin{figure}[H]
    \centering
    \includegraphics[width=1\linewidth]{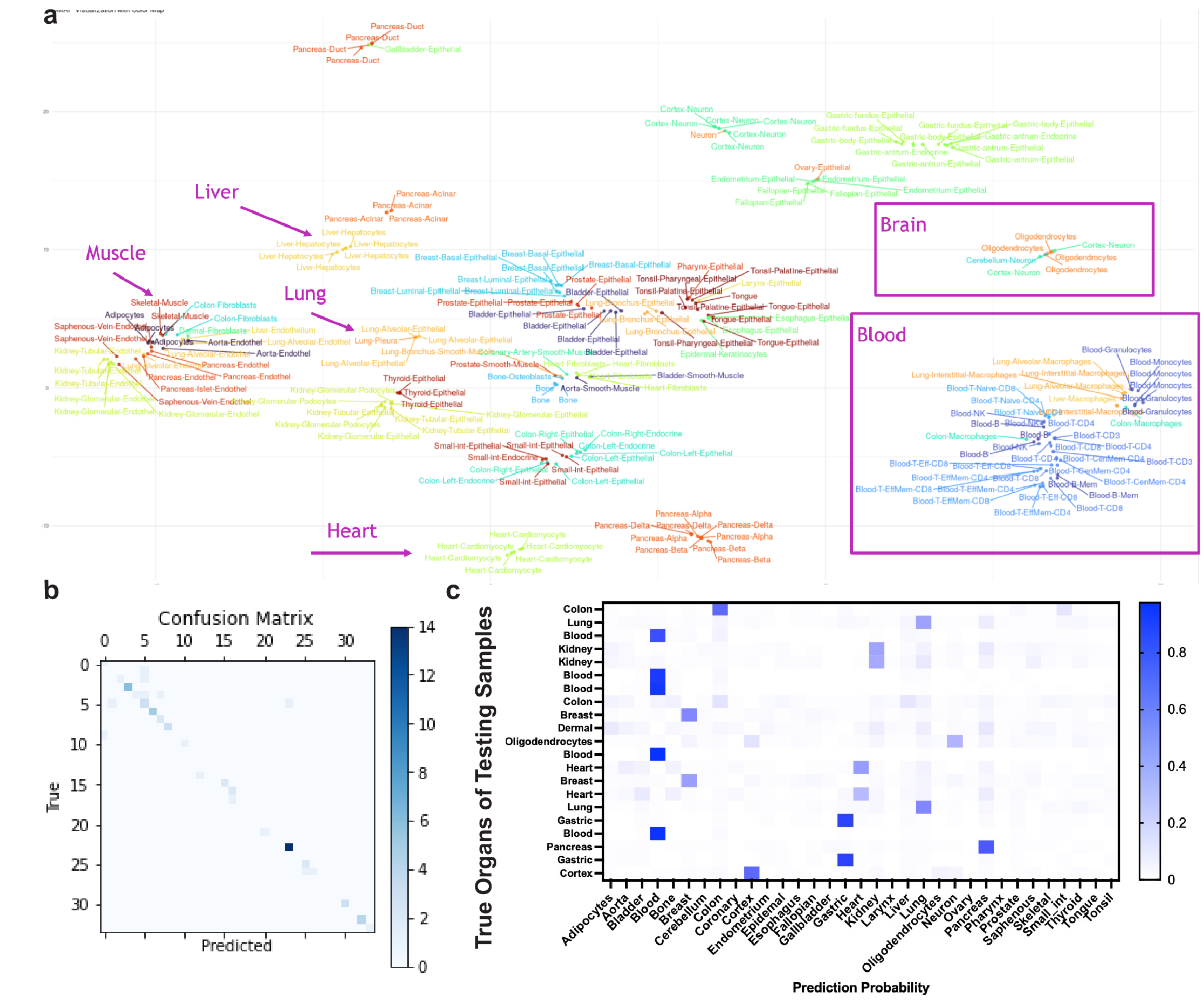}
    \caption{Characterization of DNA methylation prediction model. \textbf{(a)} The dimensional reduction of different cell types derived from various organs. \textbf{(b)} Confusion matrix of prediction from the Random Forest model. \textbf{(c)} Prediction probability of each organ labels from the Random Forest model.}
    \label{fig:enter-label}
\end{figure}
\subsection{In silico mixture validations}
To further evaluate the deconvolution capability of our DNA methylation-based classification model, we constructed an in silico cfDNA mixture experiment. In this setup, we generated synthetic samples by mixing methylation profiles from two tissues—e.g., lung, heart, or brain—at varying proportions (e.g., 70:30 and 90:10), simulating heterogeneous cfDNA derived from multiple tissue sources. The goal was to assess whether the model could correctly identify and quantify the predominant tissue signal from complex inputs, which closely resembles real-world cfDNA scenarios such as cancer, inflammation, or organ damage.\\\\
As shown in Figure 3, the Random Forest model outputted prediction probabilities across all tissue classes for each mixture. The results demonstrated that the model consistently assigned the highest probability to the tissue with the dominant proportion in the mixture (e.g., "Lung70" and "Lung90" samples showed peak probabilities for the lung class), validating its sensitivity to underlying methylation contributions. Even at lower proportions (e.g., 30\%), the model maintained detectable signal, although the prediction confidence decreased proportionally, as expected.\\\\
Importantly, the probabilistic output provides a quantitative estimate of tissue composition, not just classification. This supports the model’s utility not only for identifying tissue-of-origin but also for quantifying relative tissue contributions in mixed cfDNA samples. These findings underscore the potential of our approach for downstream applications in tissue injury profiling, transplant monitoring, and tumor burden estimation in liquid biopsy analyses.
\begin{figure}[H]
    \centering
    \includegraphics[width=0.88\linewidth]{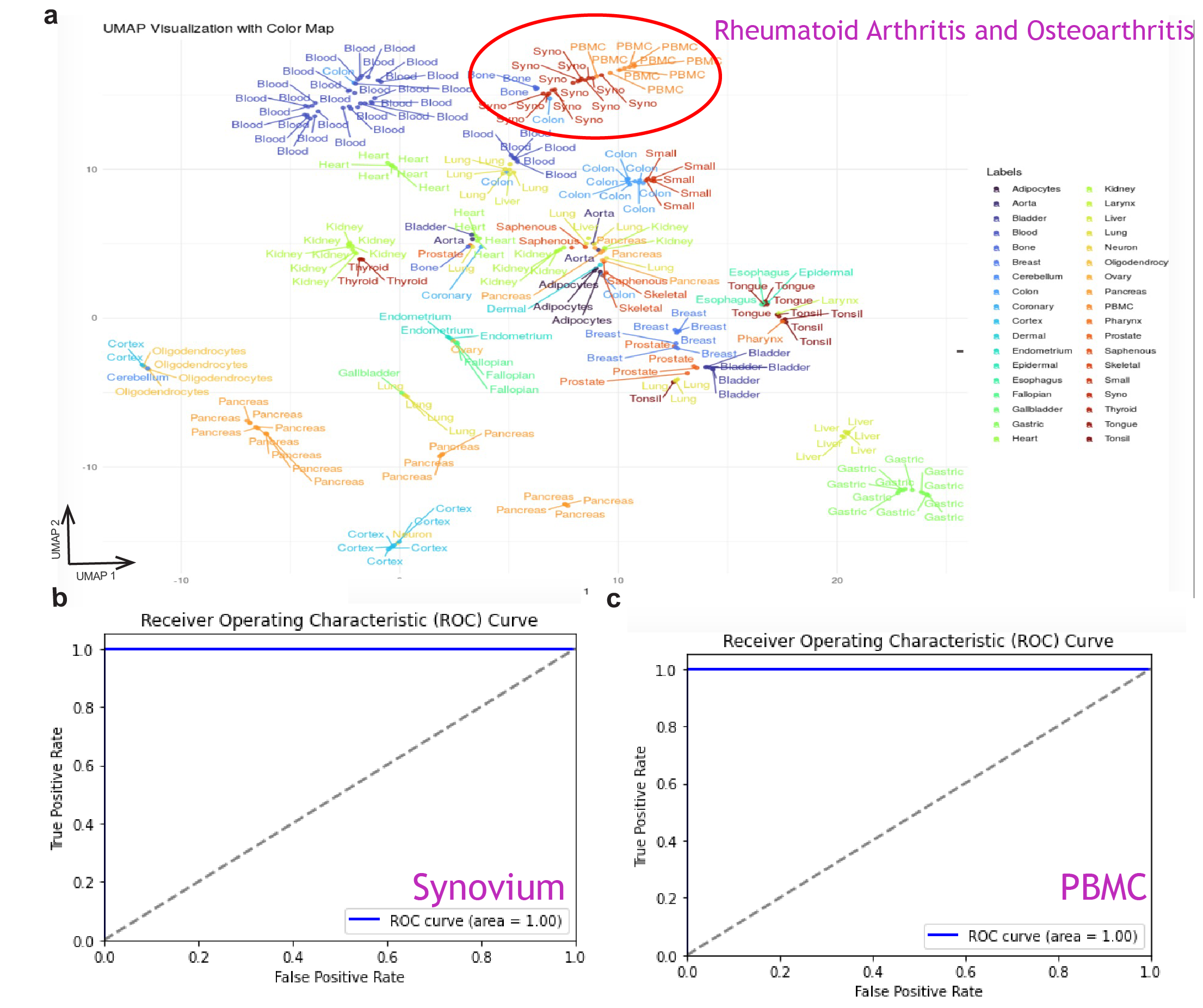}
    \caption{Characterization of DNA methylation prediction model. \textbf{(a)} The dimensional reduction of different cell types derived from various organs. \textbf{(b)} Confusion matrix of prediction from the Random Forest model. \textbf{(c)} Prediction probability of each organ labels from the Random Forest model.}
    \label{fig:enter-label}
\end{figure}

\begin{figure}[H]
    \centering
    \includegraphics[width=0.9\linewidth]{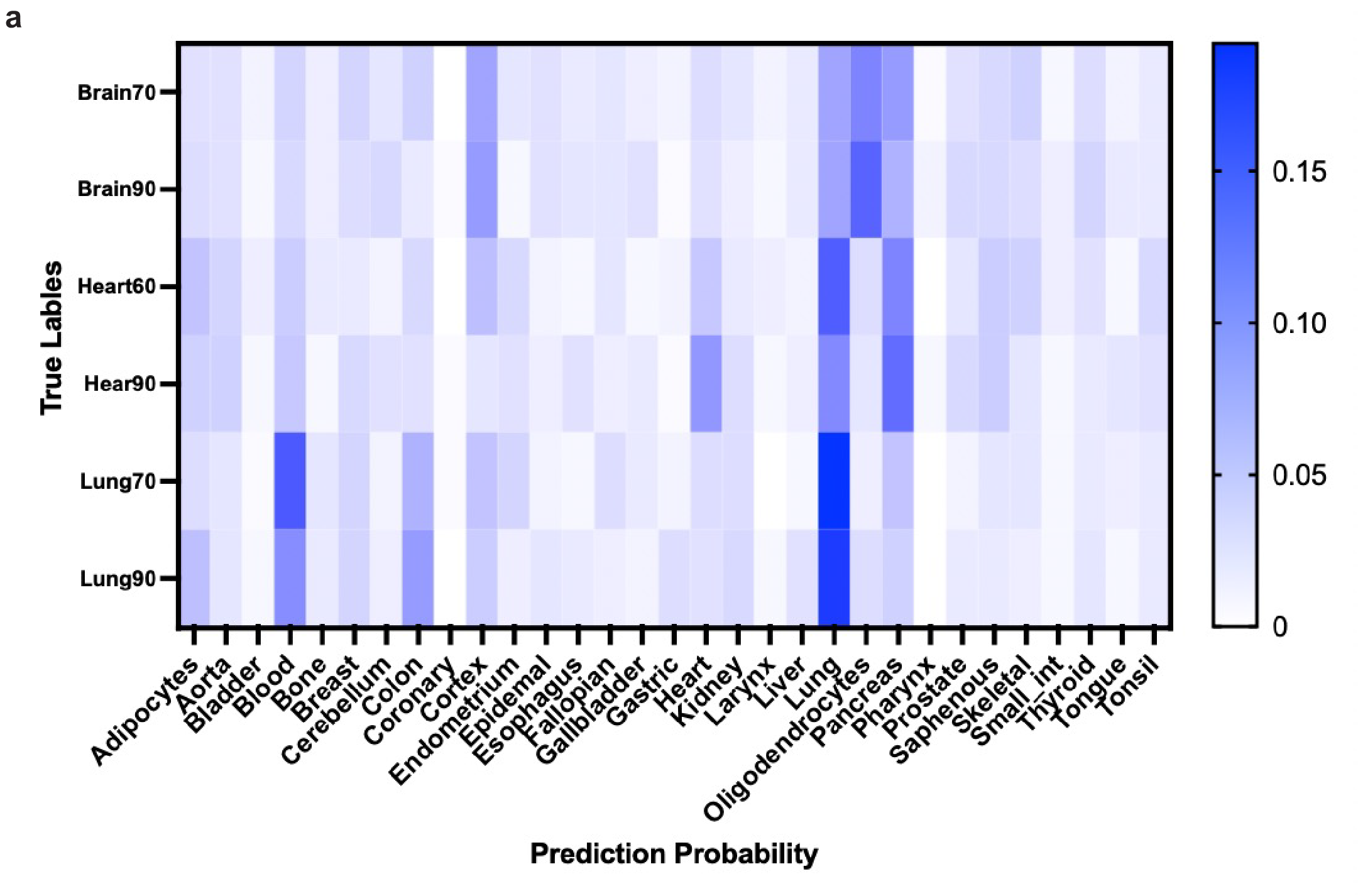}
    \caption{Characterization of DNA methylation prediction model. \textbf{(a)} The dimensional reduction of different cell types derived from various organs. \textbf{(b)} Confusion matrix of prediction from the Random Forest model. \textbf{(c)} Prediction probability of each organ labels from the Random Forest model.}
    \label{fig:enter-label}
\end{figure}
\subsection{Cross-Platform Validation of Methylation-Based Tissue Classification Using Targeted and Whole-Genome Assays}
To assess the generalizability and cross-platform robustness of our cfDNA tissue classification model, we evaluated its performance on datasets generated using both whole-genome and targeted methylation sequencing platforms. Specifically, we tested the model—originally trained on WGBS data—on methylation profiles generated by WGEM-Seq (Whole Genome Enzymatic Methyl-seq) and TEEM-Seq (Targeted Enzymatic Methyl-seq), which represent full-coverage and reduced-representation approaches, respectively. This evaluation reflects real-world scenarios in which input data may be derived from varying assay depths and resolutions, especially in clinical cfDNA workflows.\\\\
As shown in Figure 4, our model maintained tissue-level prediction capability across both platforms, with notable performance on blood and lung samples. For instance, both Blood-WGEM-Seq and Blood-TEEM-Seq samples yielded high prediction probabilities for the blood tissue class, indicating strong consistency despite platform shifts. Similarly, Lung-TEEM-Seq and Lung-WGEM-Seq samples produced elevated prediction scores for lung, confirming that organ-specific methylation signals are preserved across sequencing protocols.\\\\
Importantly, the model was also applied to cfDNA samples profiled via TEEM-Seq. While the prediction was more diffuse due to the lower resolution of targeted sequencing, we observed meaningful enrichment in tissues such as blood and bone, which are biologically plausible sources of circulating cfDNA. These results suggest that even with reduced CpG coverage, the classifier retains partial discriminatory power, supporting its application in cost-effective, targeted methylation assays.\\\\
Overall, these findings demonstrate that our machine learning model—originally trained on high-coverage WGBS data—can generalize to both genome-wide and targeted methylation inputs. This cross-platform robustness enhances its translational potential for integration into cfDNA-based diagnostic workflows that span both research and clinical environments.

\begin{figure}[H]
    \centering
    \includegraphics[width=0.8\linewidth]{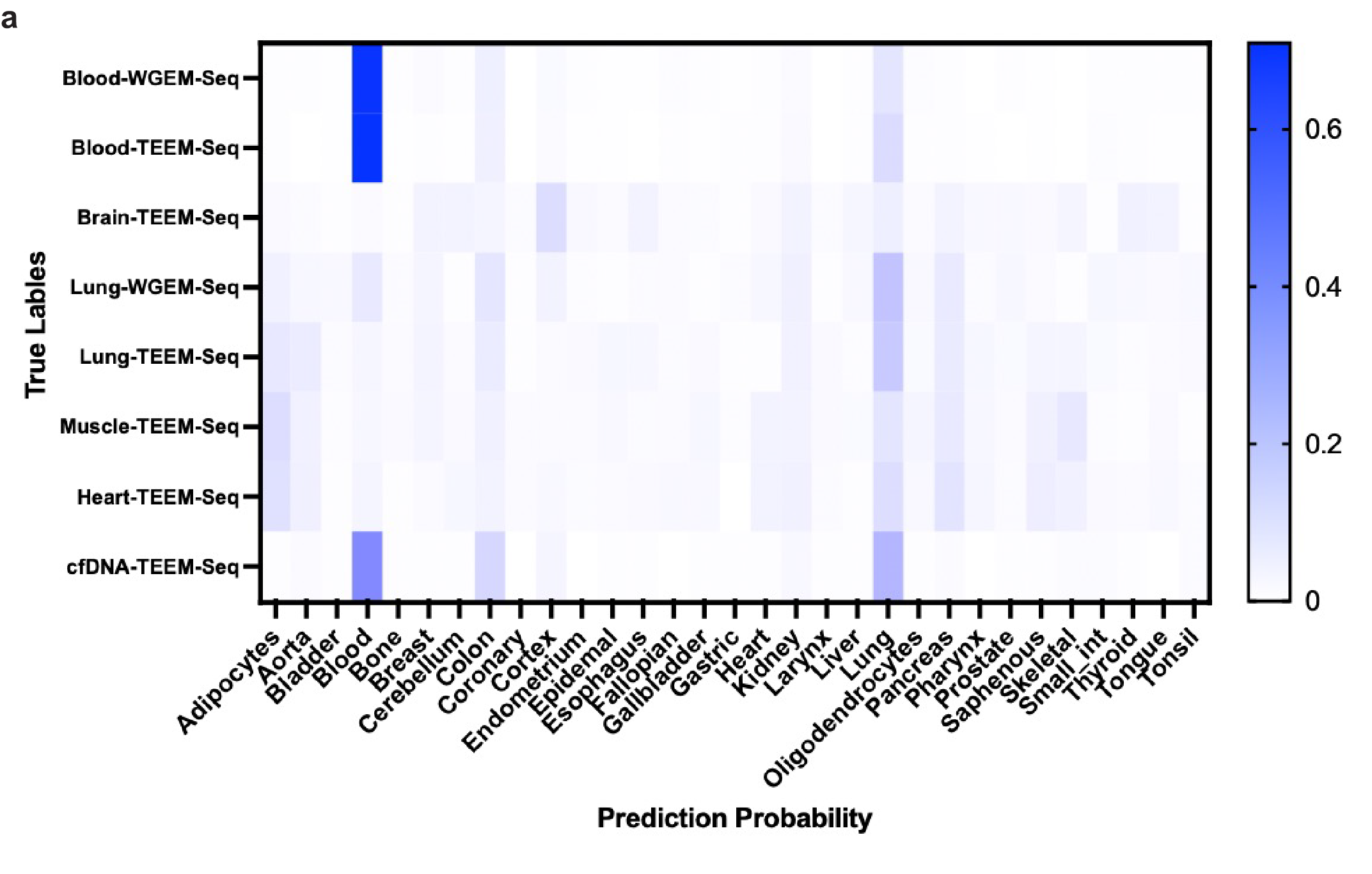}
    \caption{Cross-Platform Validation of Methylation-Based Tissue Classification Using Targeted and Whole-Genome Assays. \textbf{(a)} The heatmap of probability of Enzymatic Methylation Sequencing using targeted CpG and whole-genome.  }
    \label{fig:enter-label}
\end{figure}
\section{Conclusion}
In this study, we developed a robust, platform-agnostic machine learning framework that leverages DNA methylation signatures to classify the tissue and disease origin of cell-free DNA (cfDNA). By training on a comprehensive whole-genome bisulfite sequencing (WGBS) methylation atlas, we demonstrated that tissue-specific epigenetic patterns are sufficiently distinct to enable high-accuracy classification across a broad spectrum of human cell and organ types. Dimensionality reduction revealed strong clustering by tissue of origin, and our random forest model achieved high predictive accuracy with strong generalization performance.\\\\
Importantly, we extended our analysis beyond healthy tissues to include disease-relevant contexts, such as synovium and peripheral blood mononuclear cells (PBMCs) from patients with Rheumatoid Arthritis and Osteoarthritis. Our model accurately distinguished inflamed versus non-inflamed tissues, highlighting its potential for detecting immune-mediated pathologies. Additionally, using in silico mixtures, we validated the model’s ability to deconvolute heterogeneous cfDNA samples and quantify tissue contributions, a critical capability for real-world diagnostic applications.We further evaluated cross-platform performance using both whole-genome and targeted methylation assays (WGEM-Seq and TEEM-Seq), showing that the model retains discriminatory power across varying sequencing resolutions. This cross-platform robustness underscores the translational feasibility of our approach, especially in resource-constrained or clinical settings where targeted methylation profiling is preferred.\\\\
Together, our findings demonstrate the power of integrating tissue-specific epigenomic features with machine learning for cfDNA-based tissue classification and disease monitoring. This work lays the foundation for scalable, non-invasive diagnostics across oncology, autoimmunity, and other diseases characterized by tissue injury or immune activation. Future efforts will focus on expanding the model to include longitudinal cfDNA monitoring, integrating multi-omic features, and validating performance in clinical cohorts to support its deployment in precision medicine.

\section{Methods}
\subsection{Datasets}
205 samples, 39 cell types, 32 tissue types Methylation Atlas WGBS Dataset (GSE186458) is available at https://www.ncbi.nlm.nih.gov/geo/query/acc.cgi?acc=GSE186458. Rheumatoid Arthritis and Osteoarthritis (PBMCs and Synovium) Illumina Array Dataset (GSE164468) is available at \\ https://www.ncbi.xyz/geo/query/acc.cgi?acc=GPL21145.

\subsection{Preprocessing and Feature Engineering}
We used \texttt{wgbstools} (Loyfer et al.) to segment the genome into continuous methylation blocks with homogeneous CpG methylation levels across samples. The method employs a multichannel dynamic programming algorithm and a generative Bernoulli model, where each block $i$ in sample $k$ follows a distribution $\text{Ber}(\hat{\theta}_i^k)$. The log-likelihood of block $i$ is calculated as:

\begin{equation}
\text{score}(\text{block}_i) = ll_i = \sum_{k=1}^{K} \left( (N_C)_i^k \log(\hat{\theta}_i^k) + (N_T)_i^k \log(1 - \hat{\theta}_i^k) \right)
\end{equation}

where $(N_C)_i^k$ and $(N_T)_i^k$ are the methylated and unmethylated counts, and $\hat{\theta}_i^k$ is the Bayesian estimator:

\begin{equation}
\hat{\theta}_i^k = \frac{(N_C)_i^k + \alpha_C}{(N_C)_i^k + (N_T)_i^k + \alpha_C + \alpha_T}
\end{equation}

Regularization hyperparameters $\alpha_C, \alpha_T$ control block size. The optimal segmentation across all $N$ CpGs is computed using dynamic programming, updating a prefix score table $T[i]$ as:

\begin{equation}
T[i] = \max_{i' < i} \left\{ T[i'] + \text{score}(\text{block}_{[i'+1, \ldots, i]}) \right\}
\end{equation}

with a maximum block length of 5,000 bp. Segmentation was performed on all 205 samples using \texttt{wgbstools segment --max-bp 5000}, yielding 7,104,162 blocks. From these, 2,099,681 blocks with $\geq$ 4 CpGs were retained. For clustering, we selected the top 1\% (20,997) most variable blocks. Blocks with $\geq$ 10 CpG observations in at least two-thirds of samples were kept, and average methylation values per block were computed using \texttt{wgbstools --beta to table -c 10}. Blocks with insufficient coverage were marked as missing and imputed using \texttt{sklearn KNNImputer} (v0.24.2).

\subsection{Classification Algorithms}
To build a machine learning model for tissue-of-origin prediction from DNA methylation data, we implemented a multi-step computational pipeline. First, methylation profiles from various sequencing platforms—including Whole Genome Bisulfite Sequencing (WGBS), Illumina Methylation BeadArray, and Targeted Enzymatic Methyl-seq (EM-seq)—were processed using wgbstools to perform genome-wide segmentation. The genome was divided into approximately 80 million blocks, each representing a region of homogeneous CpG methylation, and average methylation levels were computed per block across samples to serve as input features.

Missing methylation values were imputed using the K-Nearest Neighbors (KNN) algorithm implemented in scikit-learn. We then trained and evaluated multiple classification algorithms, including KNN, Random Forest, Support Vector Machine (SVM), and Neural Network models (also using scikit-learn) to identify the tissue or organ of origin for each sample.

To ensure robust model evaluation and prevent overfitting, the dataset was randomly partitioned into 70\% training and 30\% independent testing sets. Ten-fold cross-validation was performed on the training set for model selection and tuning. Prediction performance was assessed by computing classification accuracy, defined as the number of true positive predictions (TP; predicted organ matched the true organ label) divided by the total number of predictions.

\printbibliography
\end{document}